\documentclass[aps,prl,showpacs,twocolumn,floatfix]{revtex4-1}

\usepackage{graphicx}
\usepackage{dcolumn}
\usepackage{amsmath}
\usepackage[latin1]{inputenc}

\begin{document}

\title{Are magnetite (Fe$_3$O$_4$) films on MgAl$_2$O$_4$ auxetic?}
\author{M. Ziese}
\email{ziese@physik.uni-leipzig.de}
\affiliation{Abteilung Supraleitung und Magnetismus, Universit\"at Leipzig, D-04103 Leipzig, Germany}

\date{\today}

\begin{abstract}
Magnetite (Fe$_3$O$_4$) films were fabricated on MgAl$_2$O$_4$ $(001)$ single crystal substrates
by pulsed laser deposition. In-plane and out-of-plane lattice constants were determined by X-ray diffraction.
The apparent Poisson's ratio was determined as the negative ratio of the out-of-plane to in-plane strains.
The results show that (i) the determination of Poisson's ratio by this method is only reliable for fully strained films
and (ii) Poisson's ratio $\nu_{100} \simeq 0.3$ along the $\langle 100\rangle$ direction is positive for this archetypal ferrite.
Fe$_3$O$_4$ films grown on MgAl$_2$O$_4$ $(001)$ are not auxetic.
\end{abstract}

\maketitle
Auxetic materials are substances with a negative Poisson ratio, i.e.~these expand in the transverse direction
when stretched and contract when compressed \cite{lakes1987}. This behavior is often found in foams and organic
material such as cartilage, where it is related to the microstructure and the appearance of so-called re-entrant cells \cite{lakes1991}.
Although counterintuitive, auxetic behavior does
not violate thermodynamic principles, since the non-negativity of the elastic constants restricts the possible
values for Poisson's ratio $\nu$ in isotropic materials to the range of values $-1 < \nu < 0.5$. In anisotropic media the
situation is more complicated, since Poisson's ratio depends in an intricate way on the values of the elastic moduli $C_{ij}$
as well as on the directions of the longitudinal stress and transverse strain. For cubic materials this has been studied in
detail \cite{turley1971,gunton1975,tokmakova2005,ting2005,wojciechowski2005,norris2006,paskiewicz2007}.
In case of Poisson's ratio in the $\langle 100\rangle$
direction, which is of interest here, one obtains
\begin{equation}
\nu_{100} = \frac{C_{12}}{C_{11}+C_{12}}\, .
\label{n100}
\end{equation}
With the stability criteria, see e.g. \cite{paskiewicz2007,landau}, $C_{11} > 0$, $-C_{11}/2 < C_{12} < C_{11}$ and $C_{44} > 0$,
one again obtains $-1 < \nu_{100} < 0.5$ as for isotropic materials. In general, however, Poisson's ratio can have smaller or larger
values than these for general directions in the crystal \cite{tokmakova2005,ting2005,wojciechowski2005,norris2006,paskiewicz2007},
especially for directions close to the $\langle 111\rangle$ direction \cite{norris2006}.

Recently molecular auxetic behavior in epitaxial cobalt ferrite films grown on SrTiO$_3$ $(001)$ substrates as determined
by X-ray diffraction measurements of the in-plane and out-of-plane strains was reported \cite{valant2010}. This observation was interpreted in terms of the atomic
spinel structure, but not related to the specific microstructure of the films. Since various ferrites crystallize in the spinel structure,
especially the parent compound magnetite (Fe$_3$O$_4$), it is tempting to search for auxetic behavior in magnetite films.

Magnetite films were grown by pulsed laser deposition from stoichiometric polycrystalline targets onto MgAl$_2$O$_4$ $(001)$
substrates. Deposition conditions were a substrate temperature $T_{sub}$ between 370 and 530$^\circ$C and an oxygen partial pressure between $7\times10^{-7}$~mbar
and $1.5\times10^{-5}$~mbar. An excimer laser (Lambda Physik) operating at a wavelength of 248~nm (KrF), a repetition rate of 10~Hz
and a fluence of about $1.5$~J/cm$^2$ was used for the ablation.  The thickness of the films was measured with
a Dektak surface profile measuring system. Structural characterization of the films was made by X-ray diffraction (XRD) with a Philips X'pert
system  using  Cu  K$_{\alpha}$ radiation. Both $\theta-2\theta$ scans as well as reciprocal space maps of the $(226)$ reflections were recorded.
Magnetic characterization  of  the  films  was  made  by  SQUID magnetometry (Quantum Design model MPMS-7). For further details on film growth
and characterization see \cite{bollero2005}.

\begin{figure}[t]
\includegraphics[width=0.45\textwidth]{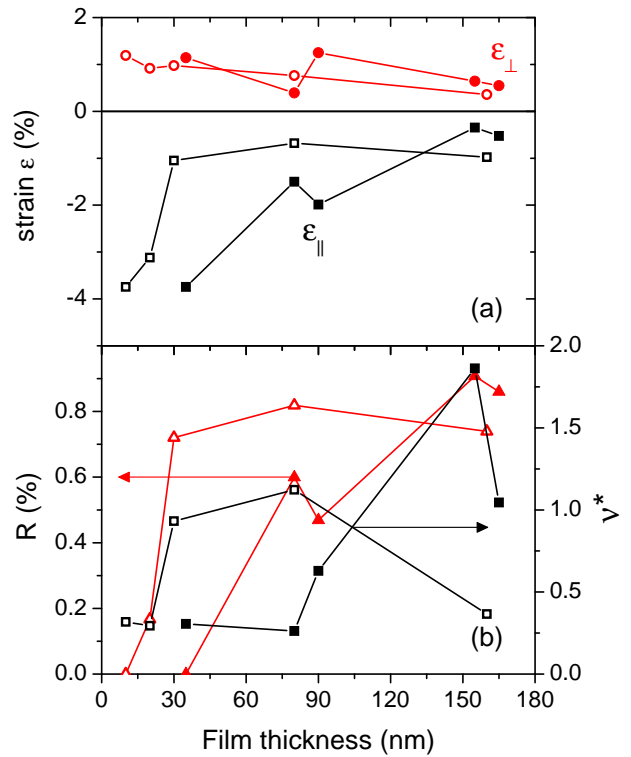}
\caption{(Color online) (a) In-plane $\epsilon_\parallel$ and out-of-plane $\epsilon_\perp$ strain as a function of film thickness.
(b) Degree of relaxation $R$ (red triangles, left axis) and apparent Poisson's ratio $\nu^*$ (black squares, right axis)
as a function of film thickness. Solid symbols: series I, open symbols: series II.}
\label{fig1}
\end{figure}

\begin{table*}
\caption{Results of the X-ray analysis. Parallel $a_\parallel$ and perpendicular $a_\perp$ lattice constants were determined from reciprocal space maps
and $\theta-2\theta$ scans, respectively. In-plane strain $\epsilon_\parallel$, out-of-plane strain $\epsilon_\perp$, degree of relaxation $R$,
volume change $\Delta V/V = (a_\parallel^2a_\perp-a_b^3)/a_b^3$ and apparent Poisson ratio $\nu^* = -\epsilon_\perp/\epsilon_\parallel$ were calculated from
the lattice constants. Film thickness $t$ and oxygen pressure $p_{O2}$ during deposition are also shown.}
\label{table1}
\begin{tabular}{llllllllll}
\hline
$t$ (nm) & $T_s$ ($^\circ$C) & $p_{O_2}$ (mbar) & $a_\parallel$ (nm) & $a_\perp$ (nm) & $R$ (\%) & $\epsilon_\parallel$ (\%) & $\epsilon_\perp$ (\%) & $\frac{\Delta V}{V}$ (\%) & $\nu^*$\\
\hline
Series I &&&&&&&&&\\
\hline
 25 & 530 &   $3\times10^{-6}$ & --     & --     &   -- & --    & --   & --    & --\\
 35 & 500 &   $3\times10^{-6}$ & 0.8084 & 0.8494 &    0 & -3.74 & 1.14 & -6.29 & 0.31\\
 80 & 450 & $1.5\times10^{-5}$ & 0.8272 & 0.8431 & 0.60 & -1.50 & 0.39 & -2.60 & 0.26\\
 90 & 450 &   $7\times10^{-7}$ & 0.8231 & 0.8503 & 0.47 & -1.99 & 1.25 & -2.74 & 0.63\\
155 & 450 &   $3\times10^{-6}$ & 0.8369 & 0.8452 & 0.91 & -0.35 & 0.64 & -0.05 & 1.86\\
165	& 370 & $1.5\times10^{-5}$ & 0.8354 & 0.8444 & 0.86 & -0.52 & 0.55 & -0.50 & 1.05\\
240	& 370 &   $3\times10^{-5}$ & --     & --     & -    & --    & --   & --    & --\\
\hline
Series II &&&&&&&&&\\
\hline
 10 & 450 & $3\times10^{-6}$ & 0.8084 & 0.8498 &    0 & -3.74 & 1.19 & -6.25 & 0.32\\
 20 & 450 & $3\times10^{-6}$ & 0.8136 & 0.8475 & 0.17 & -3.12 & 0.92 & -5.28 & 0.29\\
 30 & 450 & $3\times10^{-6}$ & 0.8310 & 0.8480 & 0.72 & -1.05 & 0.98 & -1.13 & 0.93\\
 80 & 450 & $3\times10^{-6}$ & 0.8341 & 0.8462 & 0.82 & -0.68 & 0.76 & -0.60 & 1.12\\
160 & 450 & $3\times10^{-6}$ & 0.8316 & 0.8428 & 0.74 & -0.98 & 0.36 & -1.59 & 0.37\\

\hline
\end{tabular}
\end{table*}
Two series of films with thicknesses between 25~nm and 240~nm (series I) as well as between 10~nm and 160~nm
(series II) were studied. The bulk lattice constant of the cubic phase of magnetite
at room temperature is $a_b = 0.8398$~nm, the lattice constant of MgAl$_2$O$_4$ is $a_s = 0.8084$~nm, respectively. The lattice mismatch is $-3.7$\%
and, accordingly, the films are expected to be under compressive in-plane strain. Out-of-plane lattice constants $a_\perp$ were determined
from $\theta-2\theta$ scans, in-plane lattice constants $a_\parallel$ from reciprocal space maps of the $(226)$ reflection. Out-of-plane and
in-plane strains were calculated from
\begin{eqnarray}
\epsilon_\perp &=& \frac{a_\perp-a_b}{a_b}\\
\epsilon_\parallel &=& \frac{a_\parallel-a_b}{a_b}\, ,
\end{eqnarray}
and the degree of relaxation was defined as
\begin{equation}
R = \frac{a_\parallel-a_s}{a_b-a_s}\, .
\end{equation}
The results of the X-ray analysis are presented in table~\ref{table1} and are illustrated in Fig.~\ref{fig1}.
Although for all films the volume is smaller than the bulk volume, the in-plane strain is negative and compressive,
whereas the out-of-plane strain is positive and tensile. The thinnest films of each series are fully strained, but thicker films
show considerable strain relaxation approaching almost 100\%, see Fig.~\ref{fig1}(b) (left scale). The films of series II
were grown under identical conditions for substrate temperature and oxygen partial pressure, whereas both parameters were varied
for the films of series I. This might explain that the thickness at which strain relaxation sets in, is significantly different.
However, the strain state is controlled by many factors such as fluence, growth rate and substrate quality that are not
at all times exactly reproduced.

The apparent Poisson ratio was defined as \cite{valant2010}
\begin{equation}
\nu^* = -\frac{\epsilon_\perp}{\epsilon_\parallel}\, .
\label{poisson}
\end{equation}
This is presented in table~\ref{table1} and shown in Fig.~\ref{fig1}(b) (right scale). The apparent Poisson ratio is positive for
all film thicknesses studied here. For the fully strained films a conventional value around $0.3$ is obtained, whereas in relaxed films
unphysically large values were found. This is reasonable, since the strain state in a relaxed film is certainly inhomogeneous
such that the definition of the Poisson ratio via averaged strain values is not valid. Literature values for the elastic moduli with
$C_{11}$, $C_{12} =$
270, 108~GPa \cite{baltzer1957},
268, 106~GPa \cite{li1991},
260, 148~GPa \cite{reichmann2004}, and
312, 184~GPa \cite{chicot2010}
yield Poisson ratios $\nu_{100} = 0.29$, $0.28$, $0.36$, and $0.37$. These values are in good agreement with the value obtained here for
the fully strained films.
\begin{figure}[t]
\includegraphics[width=0.45\textwidth]{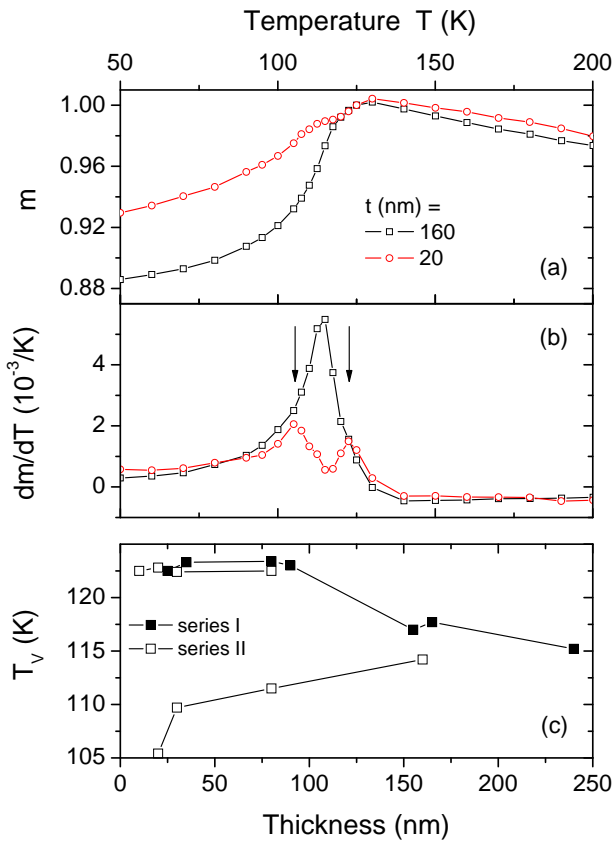}
\caption{(Color online) (a) Magnetic moment of the 20~nm and 160~nm magnetite films of series II normalized to the value at 130~K.
(b) Temperature derivative of the normalized magnetic moment. The maxima indicate the Verwey temperature $T_V$.
(c) Verwey temperature $T_V$ vs. film thickness.}
\label{fig2}
\end{figure}
One problem in the growth of oxide films lies in the stabilization of the correct oxygen stoichiometry.
Oxygen stoichiometry was indirectly assessed in the films
by the measurement of the Verwey temperature, i.e.~the transition temperature of the transformation from the high temperature cubic
into the low temperature mono\-clinic phase.
The Verwey temperature sensitively depends on the oxygen stoichiometry \cite{kakol1989}.
Fig.~\ref{fig2} shows (a) the normalized magnetic moment $m$ and (b) its derivative $dm/dT$ of the fully strained and strain relaxed
films from series II with thicknesses of 20~nm and 160~nm, respectively. The thick film showed a single
transition at the Verwey temperature of 114~K, whereas the thin film showed a double transition (most clearly seen
by the two peaks in the derivative $dm/dT$) with Verwey temperatures of 122.8 and 105.4~K. Except for the thickest film all films of series II
showed a double transition, whereas all films from series I showed a single transition. The corresponding Verwey temperatures
are shown in Fig.~\ref{fig2}(c) as a function of thickness. There are two trends: with decreasing thickness the Verwey transition
temperature of most of the films increases to a value near 123~K which is close to the bulk value of 125~K; if a double transition is
present, the lower transition temperature decreases with decreasing film thickness. The latter behavior was reported before
and was related to strain effects \cite{lindley1997,sena1997,arora2006,orna2010}. The present data indicate that -- even if the
films appear fully strained in X-ray diffractometry measurements -- some films might be in an inhomogeneous strain state with a
magnetite layer with high Verwey temperature adjacent to the substrate and another layer with lower Verwey temperature and probably
different microstructure on top of the first layer. Without further structural investigations it is impossible to determine
whether strain, microstructural effects or deoxygenation is the main factor determining the value of the Verwey
temperature and its dependence on the film thickness. Overall, however, the Verwey temperature of the thin films
is close to the bulk value indicating a nearly ideal oxygen stoichiometry.

The results presented here show that magnetite films grown on MgAl$_2$O$_4$ are not auxetic, at least not along the $\langle 100\rangle$ direction. Poisson's ratio
$\nu^* = 0.3$ determined from the in-plane and out-of-plane strain ratio of a fully strained film is in good agreement with the value derived from
measurements of the elastic moduli. Since both magnetite and cobalt ferrite crystallize in the inverse spinel structure,
these data cast doubt on the claim of molecular auxetic behavior in cobalt ferrite films grown on SrTiO$_3$ $(001)$ \cite{valant2010}.
Indeed, using the experimentally measured elastic moduli of cobalt ferrite,
$C_{11} = 257$~GPa, $C_{12} = 150$~GPa \cite{li1991}, or the calculated values $C_{11} = 240-282$~GPa, $C_{12} = 137-168$~GPa
\cite{fritsch2010}, a positive Poisson's ratio $\nu_{100} = 0.37$ is obtained that is incompatible with auxetic behavior.
A similar conclusion against auxetic behaviour was reached in \cite{iliev2011} for NiFe$_2$O$_4$ films on MgAl$_2$O$_4$.
The data presented here clearly show that Poisson's ratio can be reliably determined from X-ray measurements on fully strained films.

This work was supported by the DFG within SFB~762 ``Functionality of Oxide Interfaces''. I thank
H.~C.~Semmelhack for the X-ray diffractometry and R.~H\"ohne for the magnetization measurements.
%
%

\end{document}